\documentclass[a4,12pt]{article}
\usepackage{graphicx}
\usepackage[space]{grffile}
\usepackage{latexsym}
\usepackage{textcomp}
\usepackage{longtable}
\usepackage{tabulary}
\usepackage{booktabs,array,multirow}
\usepackage{amsfonts,amsmath,amssymb}
\usepackage{pifont}
\usepackage[round]{natbib}
\usepackage{url}
\usepackage{hyperref}
\hypersetup{colorlinks,%
                 citecolor=blue,%
                 filecolor=blue,%
                 linkcolor=blue,%
                 urlcolor=blue,%
                 pdftex}
\usepackage{fancyhdr}
\graphicspath{{./submit-FIGknotresolver/}}

\title{Directed graphs and curve matching to track self-intersecting filaments in microscopy}
\author{Dhruv Khatri$^1$, Shivani A. Yadav$^1$ and Chaitanya A. Athale$^{1*}$,\\ 1: Div. of Biology, IISER Pune, Dr. Homi Bhabha Road, Pashan,\\ Pune 411008, India}
\date{}

\fancypagestyle{plain}{%
   \fancyhead{} 
}

\begin{document}
\begin{center}
{\Large {\bf KnotResolver: Tracking self-intersecting filaments in microscopy using directed graphs}}
\end{center}
{\it Dhruv Khatri$^1$, Shivani A. Yadav$^1$ and Chaitanya A. Athale$^{1*}$,\\ 1: Div. of Biology, IISER Pune, Dr. Homi Bhabha Road, Pashan,\\ Pune 411008, India}

\section*{Summary}
Khatri et al. have proceeded to address the limitations of existing computational tools to track self-intersecting single cytoskeletal filament dynamics in microscopy image time-series. Their approach, `KnotResolver’ combines directed graphs and contour warping to result in robust and accurate results.

\section*{Abstract}
Quantification of microscopy time-series of {\it in vitro} reconstituted motor driven microtubule (MT) transport in `gliding assays' is typically performed using computational object tracking tools. However, these are limited to non-intersecting and rod-like filaments. Here, we describe a novel computational image-analysis pipeline, KnotResolver, to track image time-series of highly curved self-intersecting looped filaments (knots) by resolving cross-overs. The code integrates filament segmentation and cross-over or `knot' identification based on directed graph representation, where nodes represent cross-overs and edges represent the path connecting them. The graphs are mapped back to contours and the distance to a reference minimized. We demonstrate the utility of the tool by segmentation and tracking MTs from experiments with dynein-driven wave like filament looping. The accuracy of contour detection is sub-pixel accuracy, and Dice scores indicate a robustness to noise, better than currently used tools. Thus KnotResolver overcomes multiple limitations of widely used tools in microscopy of cytoskeletal filament-like structures. 

\section*{Introduction}
Microtubules (MTs) are an essential component of the cell cytoskeleton and their interaction with molecular motors serves to generate forces required for organelle positioning and the assembly of supra-molecular structures like the mitotic spindles, cilia and flagella \citep{Woolley:2000vh,Guerin:2010vh}. The collective properties of MT growth and interactions result in complex networks {\it in vivo}. However, quantifying the mechanical properties in such an intracellular environment has proven to be difficult due to the crowded nature of the intracellular space \citep{ellis2001macromolecular}. As a result, collective the mechanical properties of both MTs and motors have been extensively studied {\it in vitro}. One prominent method used to examine collective transport by molecular motors is by immobilizing motors on the surface and allowing filaments to bind to them, which results in transport of MTs, referred to as `gliding assays' \citep{Howard:1989aa,Nitzsche:2010aa,Sumino:2012aa}. Typically in such microtubule `gliding assays' with short filaments, i.e. of lengths 5 $\mu$m or less, are observed almost as rod-like objects in microscopy in presence either kind of motor- kinesin \citep{Scharrel:2014aa} or dynein \citep{Monzon:2018aa,jain2019}. Actin filament motility driven by myosin however shows highly curved and bent structures at similar length scales \citep{Kron:1986aa,Kron:1992aa,Ishijima:1991aa}. This is consistent with the reported persistence length of MTs {\it in vitro} at thermal equilibrium of  the scale of millimeters while actin has an average persistence length of the order of $\sim$10 $\mu$m \citep{gittes1993}. Additionally highly curved filaments are seen in `gliding assays' either due to torsional forces exerted by myosin resulting in filaments `twirling' \citep{Beausang:2008aa}, and MTs when held in optical tweezers \citep{kurachi1995buckling} or pinned at one end due to buckling \citep{gittes1996}. 
In recent work, complex curved structures of bundled MTs have been observed in bottom-up reconstitution of axonemal bending \citep{guido2022}, as well as plus-end clamped MTs in a dynein `gliding assay' that undergo cyclical wave-like oscillations and self-intersections \citep{Yadav:2024aa}. Thus tracking curved and self-intersecting filaments, i.e. knots, from time series are of wider relevance for quantification of collective motor driven filament transport studies.

Quantifying such `gliding assays' typically involves detecting and tracking the movement of individual filaments in microscopy time-series. Some of the widely used tools include a MATLAB based tool FIESTA, that was shown to track MTs with sub-micron precision \citep{ruhnow2011tracking}, an active countour based approach applied to track actin filaments, JFilament that works as a FIJI plugin \citep{li2009,smith2010segmentation} and MTrack, a versatile FIJI plugin \citep{kapoor2019mtrack}. While these tools are optimized for a general `gliding assay', they assume non-intersecting filaments that are straight over the time-series, assumptions that are violated in either complex geometries or crowded environments.

Multiple algorithmic approaches to resolving networks from microscopy have been developed including a general algorithm for the decomposition of filamentous networks (DeFiNe) by weighted, undirected graph representation and roughness minimization that can resolve simulated, cellular and astronomical images of meshes of filaments \citep{breuer2015define}. At a whole cell scale NeuriteTracer has been used to automatically detect neuronal extensions based on low-level segmentation and skeletonization \citep{Pool:2008aa}, which has been integrated with a graph theoretical representation in simple neurite tracer (SNT) \citep{Arshadi:2021aa}. At an organismal scale, images of \textit{Caenorhabditis elegans} worms at high density were resolved by combining skeletonization with a graph-theoretical representation of worm contours as edges and adjacencies as nodes  \citep{raviv2010morphology}. Representing networks as open curves using B-spline surface representation of filaments optimized by gradient-descent to the image, combined with image-formation model to detect and segment unknown numbers of filaments \citep{xiao2016automatic}. Thus, across problem-types the application of graph representations have been useful to resolve filaments in complex geometries. However these have, to our knowledge, not yet been applied to single filaments with cross-overs in combination with tracking. 

In this study, we describe a novel approach to segment and track single microtubule filaments in time-series data that undergo self-intersections or knots, that we refer to as KnotResolver. Our approach combines active contours seeded by an initial guess based on an intensity threshold and geometry, the representation of the intersecting segments as elements of a directed graph and distance minimization between time-frames by comparing curves predicted by the directed graph with an initial reference that is not intersecting. We demonstrate the utility of our approach by applying it to resolve knots in both {\it in silico} and {\it in vitro} oscillations of clamped MTs described in a recent study \citep{Yadav:2024aa}.  
We demonstrate the improvements in tracking by comparing KnotResolver to existing filament tracking tools.

\section*{Results}
\subsection*{Filament looping and cross-over detection: limits of existing tools}
\begin{figure}[ht!]
\centering
\includegraphics[width=0.9\textwidth]{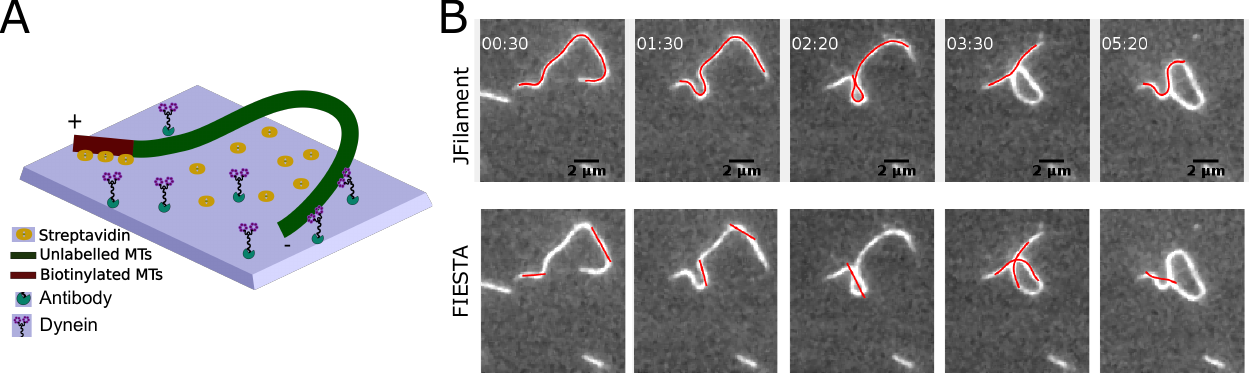}
\caption{\textbf{{\it In vitro} assay of filament oscillations and errors in detection using Gaussian fitting and active contours.} (A) The schematic depicts the experimental setup of plus-end clamped MTs based on biotin-streptavidin chemistry with dynein motors immobilized by antibodies in a `gliding assay' that drives MT bending as described previously \citep{Yadav:2024aa}. (B) Representative frames from a microscopy time-series of the oscillations of rhodamine-labelled MTs (gray) oscillating over 5 min 20 s were segmented detected segments overlaid (red lines) based on either {\it (top)} an approach using open, active contours in JFilament \citep{smith2010segmentation}  or {\it (bottom)} threshold and Gaussian fitting approach using FIESTA \citep{ruhnow2011tracking}. Scalebar: 2 $\mu$m, Time: mm:ss.}
\label{fig:clampTrkFiesta}
\end{figure}

We have recently demonstrated microtubules can undergo waves of high curvature and oscillate in a modified {\it vitro} `gliding assay', where plus-ends of MTs were clamped to the glass substrate while a surface-immobilized minus-end directed motor, dynein generated the transport force\citep{Yadav:2024aa} (Fig. \ref{fig:clampTrkFiesta}A). 
The filaments undergo wave-like millihertz (slow) oscillations in a motor-density and filament length dependent manner. In order to quantify filament contours from such image time-series data, we applied multiple standard image-analysis tools for segmenting filaments in `gliding assays' of which we describe the result of using the two more widely used: JFilament and FIESTA. JFilament \citep{li2009,smith2010segmentation} based on a robust active-contour approach, resulted in filament detections, only when they were not looped, but once filaments underwent self-intersections the method took what appeared to be the shortest path (Fig. \ref{fig:clampTrkFiesta}B{\it (top)}). This relates to the nature of the active contour method used.  FIESTA \citep{ruhnow2011tracking} on the other hand, is based on Gaussian profile fitting to improve threshold-based segmentation with filament tracking broadly based on distance minimization. The outcome of using the same dataset, resulted in detection only of the straight segments of filaments, with highly curved regions of the same filament fragmented into multiple straight segments (Fig. \ref{fig:clampTrkFiesta}B{\it (bottom)}). While tuning multiple parameters of the code did change the results slightly, it did not lead to a qualitative improvement, due perhaps to the inherent assumption of straight rod geometry built into the method. Thus we proceeded to address both these limitations of the existing approaches- straightness assumption and path length optimization- to develop a novel approach. 

\subsection*{Resolving self-intersecting `knots' in filament contours by minimizing paths using directed graphs}
\begin{figure}[ht!]
\centering
\includegraphics[width=0.9\columnwidth]{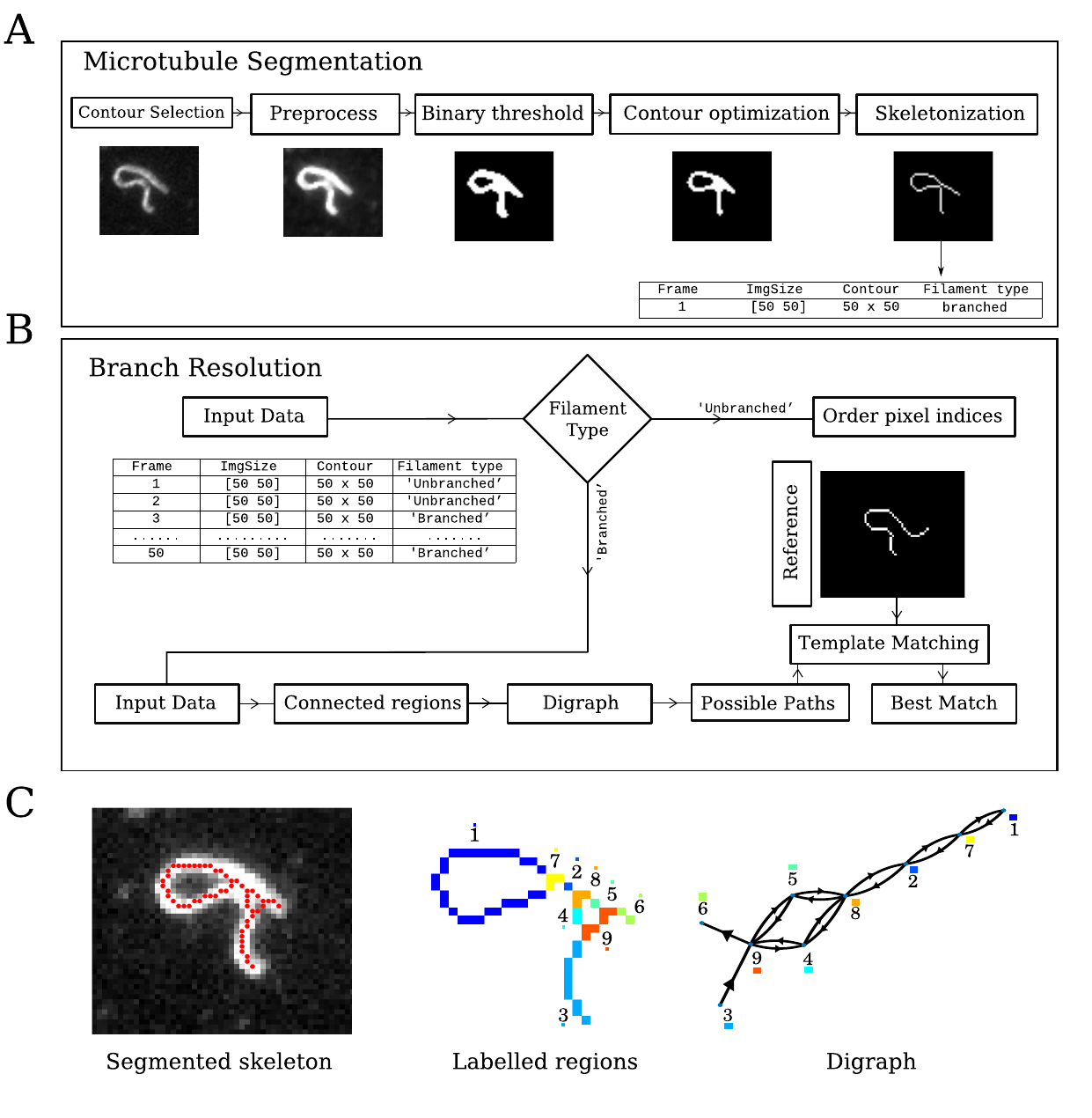}
\caption{\textbf{Algorithmic flow diagram of graph-based resolution of self-intersecting filament image time-series.} (A) {\it Segmentation and branch finding:} In a frame-wise manner images are binarized by thresholds. 
These binary objects are used to seed active contours to smooth the curve. 
The contour is skeletonized to 1 pixel width using medial axis transform, and pixels with 4 or more neighbors are labelled as branched, suggesting a self-intersection. 
(B) {\it Branch resolution:} The skeleton with sequential pixels is used as an input labelled as either `branched' or `unbranched'. Branched contours are mapped to a directed graph (digraph) with branch points as vertices connected to each other by segments of the contour as edges. Paths are calculated using a two-step coarse and refined approach to reduce complexity. Graph-paths are mapped back to possible contours, whose distance is then minimized by comparison to the last unbranched contour (template), to find the best match. (C) ({\it left to right}) A representative grayscale image (gray) segmented resulting in a skeleton contour (red), that is classified into regions (colors, numbers) that connect branch points. These regions are mapped to edges and branches to vertices, with both directions possible based on a user-input start point, to produce a directed graph.}
\label{fig:algo}
\end{figure}

The `KnotResolver' pipeline consists of the following steps: (i) {\it Segmentation} used to extract skeletons of a single microtubules (Fig. \ref{fig:algo}A). Segmentation parameters are interactively set by visually inspecting the output with the optimal values saved in a CSV file. (ii) {\it Branch identification} is called after the segmentation optimization script, that identifies filament self-intersections (Fig. \ref{fig:algo}B). {\it (iii) Mapping `knots' to graph:} converts the elements into a graph representation, {\it (iv) Template matching} to reconstruct the correct filament geometry by path minimization to previous time-frames (Fig. \ref{fig:algo}B,C). 

{\it (i) Segmentation:} Images were first pre-processed to enhance contrast by saturating the top and bottom 1\% of intensity values and median filtering using a 3x3 kernel to smooth the intensity fluctuations. The pre-processed image stack was converted to binary (black-white) by Otsu's method of finding the threshold\citep{otsu1979threshold} and the segmented image with filament contours used as the initial seed for an energy-based active contour model based on Chan and Vese's method \citep{chan2001active}. This improves the segmentation by optimizing for the filament boundary. Finally, the optimized binary contour is skeletonized using the medial axis transform \citep{lee1994building} to obtain one-pixel width filament contours, which are stored in a structured array for all frames in the image stack (Fig. \ref{fig:algo}A,C{(left)}). 

{\it (ii) Branch identification:} Skeletons obtained in the previous step are tested for branch points in a neighborhood of 8 (Moore neighborhood), by testing whether the summation of individual pixel values ($I_{nb}$ = either 0 or 1) satisfies $\Sigma_{nb}^N(I_{nb}) \geq 4$. The central pixel ($nb=5$) has a value 1, by virtue of being on the skeleton. 
In cases where there are no branch points in a segmented skeleton, the correct orientation of the filament indices is determined using the geodesic distance algorithm \citep{soille1999morphological} implemented using the {\it bwdistgeodesic} function in MATLAB (Mathworks Inc., Natick, MA, USA). Filament end points are provided as a seed to the function. When the skeleton is identified as being  a `knot', i.e. with branches, we proceed to use a graph representation, as described in the following section. 


{\it (iii) Mapping `knots' to graphs:} A skeleton containing branch points is represented as a directed graph $G = (V,E)$ consisting of vertices \textit{V} and edges \textit{E}, with the direction given by a chosen start point. They are labelled as either branch ($V_b$) or filament segment ($V_f$) (Fig. \ref{fig:algo}B). A branch vertex is identified by connected pixels that satisfy the branch-point criterion, while the remaining regions are labelled as filament vertices. 
Edges \textit{E} are the connectivity between unique label groups, forming connecting pairs of $V_b$ and $V_f$ vertices (Fig. \ref{fig:algo}C). All edges are bidirectional, i.e they can be $E(V_b, V_f)$ or $E(V_f, V_b)$, enabling traversal in either direction, except for start and end vertices where traversal is limited to a single direction. 
Instances where a filament merges with itself are represented as loops, in the graph, i.e. where heads and tails coincide \citep{Bang-Jensen:2001aa}, for example for the pair of edges: E(1,7) (Fig. \ref{fig:algo}C). 

\begin{figure}[ht!]
\centering
\includegraphics[width=0.75\linewidth]{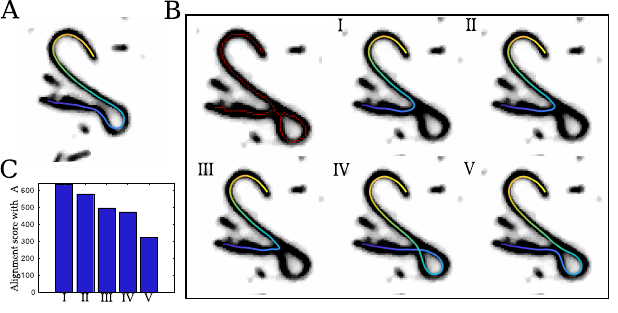}
\caption{\textbf{Resolve the knot by minimizing distances between template and paths generated by the graph.} (A) A representative image of a bend microtubule at time {\it t}, when it does not have any knots (unbranched) with the segmented contour. Colors: position index along the contour, blue: start, yellow: end. (B) The microtubule in frame {\it t+$\delta$t} showing a knot (branched) as seen from the contour (red line). (I-V) The possible paths from the start to the other end are overlaid as contours with the position-index in color. Blue: start, yellow: end. $\delta$t: interval between frames.  (C) The alignment score based on dynamic time wrapping for all paths is calculated by comparing with the unbranched path from the previous time-frame, and the path with the lowest score is the optimal.}
\label{fig:allpaths}
\end{figure}

{\it (iv) Template matching:} A branched skeleton thus represented as a graph $G = ({V_{i}, E_{i,j}})$, where $V_{i}$ denotes the set of vertices and $E_{i,j}$ denotes the set of bidirectional edges where $i$ and $j$ are indices of branched or filament vertices. The graph structure is determined by the number of unique branch points, which either indicate intersections with other filaments or an instance where the filament forms a loop. We generate all possible path combinations between the start and endpoint by traversing every edge connecting vertices $V_{f}$ (filament vertex) and $V_{b}$ (branch vertex). For graphs that lack an end point, the path ends at the last visited node, after all nodes are visited once. Since the number of paths can increase exponentially with the number of nodes we optimized the number of possibilities generated at each step to capture filament paths correctly, based on comparison to the `ground truth', obtained from manual analysis. 
Knot resolution is based on minimizing the difference to a resolved skeleton, which is the template (Fig. \ref{fig:allpaths}A). 
We have implemented a multi-step approach: (a) coarse identification of key vertices using euclidean distance matching, (b) path generation to output possible paths between the start and end point and (c) template matching of possible paths to the previous resolved contour. 
(a) {\it Identifying key vertices:} Euclidean distance scores are of vertices compared to the template helps identify connections with minimal distance to a reference- to identify the three vertices with minimal distance to the template ($V_{m=1 to 3}$). These vertices are used to limit the search and significantly reduce the number of paths generated. (b) {\it Path generation:} Multiple paths are generated between a start node and a defined end vertex (Fig. \ref{fig:allpaths}B). The first combination of path is generated by taking into account the template found in the previous step. The remaining paths are generated by finding number of combinations between the fixed start point and all of the remaining vertices sequentially. The number of paths generated at every combination is limited to 10, to avoid exponential growth of possibilities. (c) {\it Template matching:} All possible paths are converted to contours and a similarity score is assigned using dynamic score warping, DTW \citep{sakoe-chiba1978}. The score is calculated as the cumulative cost function $D(i,j)$ between a selected contour $S(x_{i},y_{i})$ and $R(x_{j}, y_{j})$ given by: 
\begin{equation}
D(i, j) = \left| S(x_i, y_i) - R(x_j, y_j) \right| + \min \left( D(i-1, j-1), D(i-1, j), D(i, j-1) \right)
\end{equation}
where $i$ is the reference index and $j$ the template index. This cost is compared among the possible paths, and the path with the minimal cost is selected (Fig. \ref{fig:allpaths}C). Once a match is found, pixel indices are iteratively ordered, along the best found path using the geodesic distance \citep{paliwal1982modification}. 

\subsection*{KnotResolver has sub-pixel filament contour detection accuracy and is robust to image `noise'}
\begin{figure}[ht!]
\centering
\includegraphics[width=0.85\linewidth]{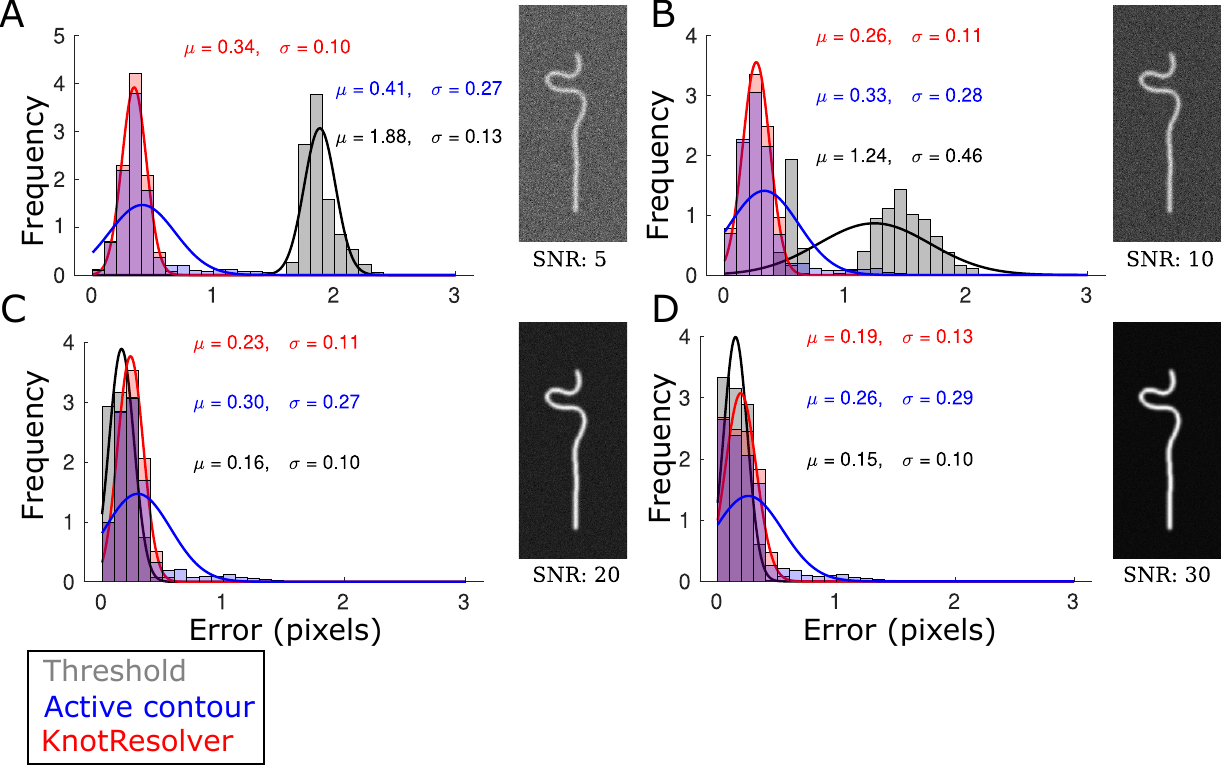}
\caption{{\bf Simulated filament images with increasing SNR used to measure position detection error comparing segmentation methods}. \textbf{(A)} Segmentation accuracy of compared methods at different signal-to-noise ratios (SNR). Simulated filaments were segmented at varying SNR levels using three different methods: the Threshold method (black), the Active contour method (blue), and our proposed method (red). The total error in pixels for each segmented pixel with respect to the ground truth is plotted as a histogram of probability density function. The inset displays the average and sigma of a Gaussian fit to each histogram \textit{(n = 2880)}. A representative image of an input filament at the specified SNR is shown adjacent to each histogram for visual reference.}
\label{fig:segError}
\end{figure}

In order to asses the segmentation accuracy of our method, we created images of simulated time series of filaments that undergo crossovers based on plus-ends clamped and minus-end directed motors in a modified `gliding assay', as described previously \citep{Yadav:2024aa} and briefly summarized in the Materials and Methods section. These simulated MT coordinates in time-series were plotted as binary skeletons to provide the `ground truth' and processed to resemble experimental images of filaments by dilation with a diamond-shaped structuring element of size 2 and convolution with a Gaussian point spread function (size: 3, $\sigma$: 2) centered around the central pixel. Salt-and-pepper noise of different intensities was added and the signal-to-noise ratios (SNR) calculated. The error in position detection is calculated as the minimal distance between the segmented skeleton and the ground truth pixel indices. 
`KnotResolver' was then used to segment the images of bent filaments, and compared to a simple intensity threshold and active-contour based approach, when the signal to noise ratio (SNR) increased from 5 (Fig. \ref{fig:segError}A), 10 (Fig. \ref{fig:segError}B), 20 (Fig. \ref{fig:segError}C) to 30 (Fig. \ref{fig:segError}D). We find the mean pixel error in position detection of `KnotResolver' and active-contour methods remains sub-pixel for the range of SNR as seen by the Gaussian peak from fitting ($\mu$: mean, $\sigma$: std. dev.), but simple threshold-based methods show errors exceeding few pixels for SNR 5 and 10. In image-frames where the filament contour self-intersect, only KnotResolver can correctly identify contours, with both segmentation and active-contour methods failing (Fig. \ref{fig:Dice}A). 
`KnotResolver' remains robust to noise for SNR ranging from 30 to 5 but threshold-based methods and active contours are sensitive, as quantified by the Dice Score as described in detail in the Methods section (Fig. \ref{fig:Dice}B). Threshold-based methods as highlighted are sensitive to low SNR while active contour methods are sensitive to rapidly changing geometry that throws off the initial seed and leads to filament shrinkage. The combination of these two methods, implement in KnotResolver is robust to both low SNR and rapidly changing filament positions which is key for tracking single filaments showing sustained oscillations. 

In order to identify possible limitations of our method and solve them, we tested the effect of multiple self-intersections that arise in simulations with increasing filament length and motor density and find Dice score decreasing after $\sim$ 60 frames in microtubules with lengths greater than 20 $\mu$m (Fig \ref{fig:sim}A). As a measure of the geometric shape, we also use the Fr\'echet distance score between segmentation result and ground truth and find a comparable increase in distance indicating drop in accuracy of segmentation for long MTs at high densities (Fig \ref{fig:sim}B). To address this limitation, we have implemented a manual restart, which can be used when the code encounters problematic frames using user-inout to interactively map the path in the graph structure and correct the errors (Fig. \ref{fig:sim}C). 

In summary while sub-pixel accuracy can be achieved by other methods alternative to the proposed KnotResolver pipeline, 
the higher Dice score in cases of self-intersections demonstrates both high accuracy and robustness to noise. 

\subsection*{KnotResolver outperforms existing methods for microtubule filament tracking}
\begin{figure}[ht!]
\centering
\includegraphics{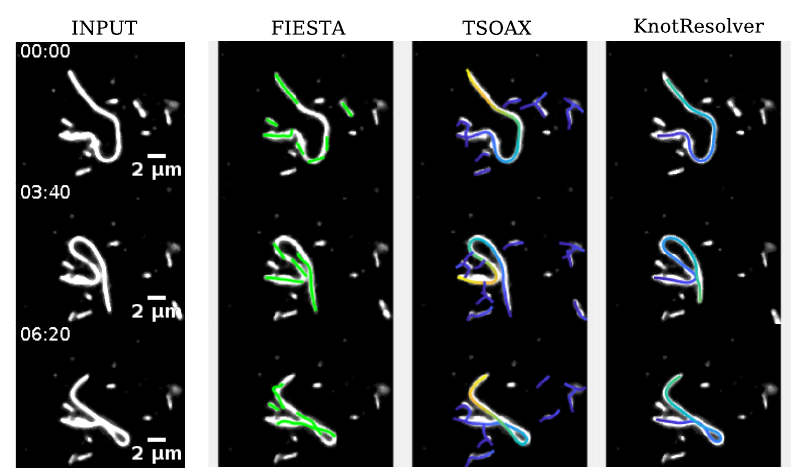}
\caption{\textbf{Comparing KnotResolver to FIESTA and TSOAX.} {(INPUT)} A representative time-series of labelled MTs undergoing bending and buckling forming a knot, was used as input and tracked using FIESTA \citep{ruhnow2011tracking}, TSOAX \citep{xu2019automated} and KnotResolver. Parameters used for (i) FIESTA were FWHM = 1000 nm, proportion curved filaments = 25\%, minimum track length = 1 frame, maximum break = 4, (ii) for TSOAX default parameters were used 
with manual input of the minimum and foreground intensity value and (iii) KnotResovler were intensity threshold = 0.46, number of iterations for active contours =10, contraction bias = 0.4 and smoothing factor for active contours = 0.2, with no manual corrections. Scale bar: 2 $\mu$m. Time: mm:ss.}
\label{fig:softwareCompare}
\end{figure}

\begin{figure}[ht]
\centering
\includegraphics[width=0.8\textwidth]{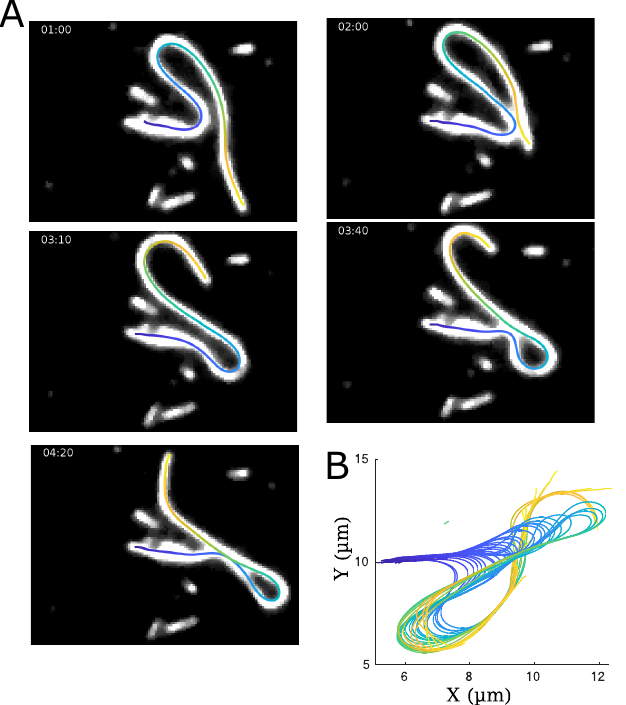}
\caption{{\bf Representative time-series of filament with tracked contours.} (A) The fluorescence time-series of bent and crossing over filaments are overlaid contour segmented using proposed pipeline. The skeleton is color coded along its length (form tip to end). Time: mm:ss. (B) The filament contours from successive frames are projected showing the traveling wave of curvature.}
\label{fig:knotTrk}
\end{figure}

\begin{figure}[ht!]
\centering
\includegraphics[width=0.85\linewidth]{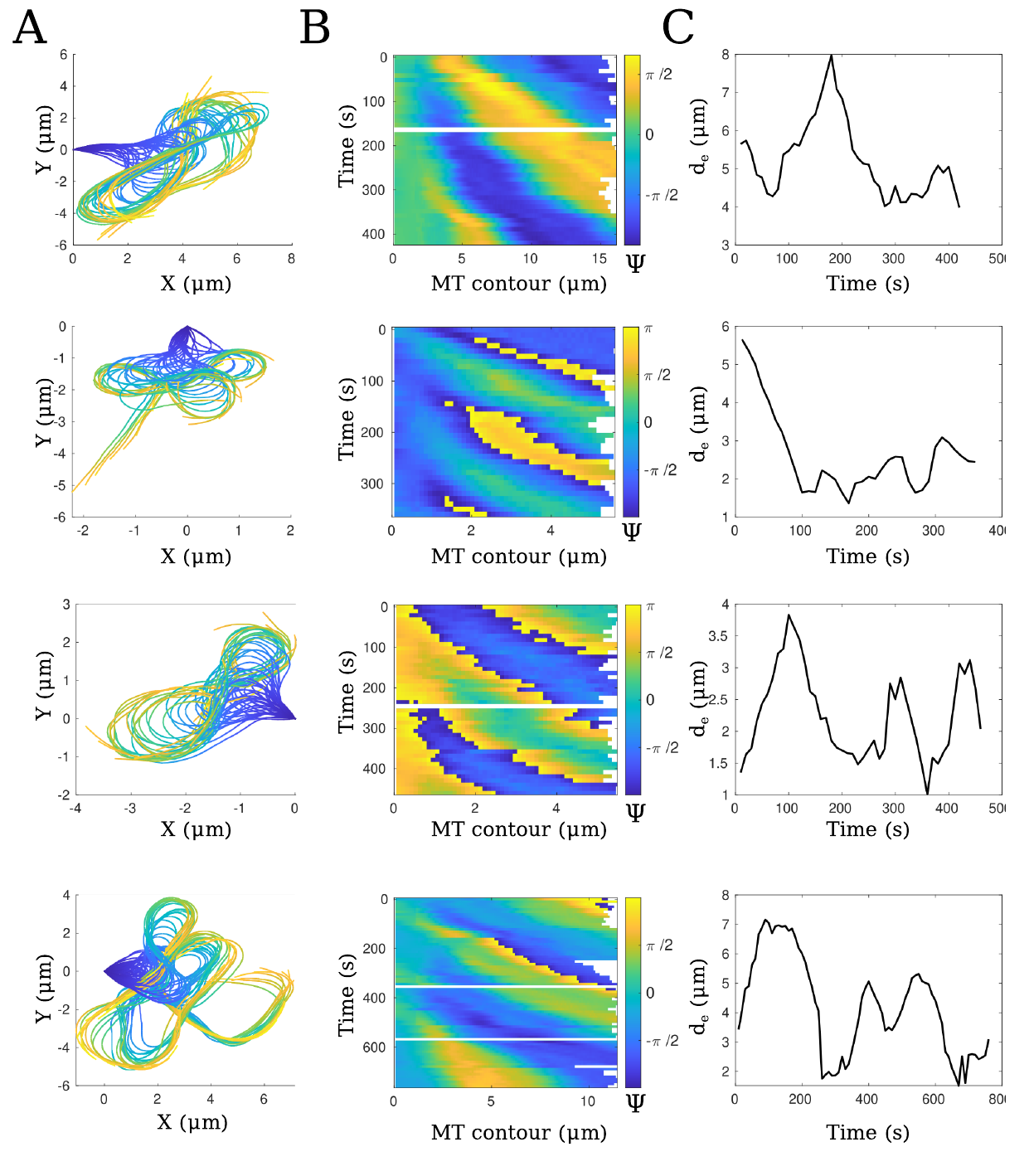}
\caption{{\bf Contour analysis in time of representative experimental filaments.} Four filaments showing beating mobility patterns are shown. \textbf{(A)} The skeletal line in the montage is color coded along its length (form tip to end). \textbf{(B)} 2D plot of tangent angle along the microtubule length (\textit{$\mu$m}) is shown. \textbf{(C)} End to end distances along the time series is shown.} 
\label{fig:Analysis}
\end{figure}

We compare of our approach to resolving self-intersecting filaments in terms of segmentation and tracking to two widely used computational tools - FIESTA \citep{ruhnow2011tracking} and TSOAX \citep{xu2019automated}. FIESTA is based on intensity thresholding and Gaussian fitting of contours to achieve sub-pixel localization accuracy. Filament intersections are semi-automatically handled with cross point identification followed by linking based on a minimum angle criterion. TSOAX combines stretching open active contours based software (SOAX) \citep{xu2015soax} with tracking by automatic initialization of multiple open curves during the detection stage that elongate and stop at filament intersections and tips, and curve intersections being addressed by identifying T-junctions linked through a temporal local matching step. We use an experimental time-series from highly curved MT filaments in a clamped gliding assay that show changes from straight through curved to self-intersecting filaments (Fig. \ref{fig:softwareCompare}{\it (input)}). The output from FIESTA demonstrates only straight segments are detected, while TSOAX while able (as expected) to identify curved filaments, results in an incorrect tracing of the filament contour at cross-over. Only KnotResolver correctly identifies curved sections and maintains the identity of the segments at cross-over. 
The entire experimental time-series can be resolved using KnotResolver and results in a smooth contour (Fig. \ref{fig:knotTrk}A, Video \ref{SV1}) which when overlaid provides time-resolved data of filament dynamics, as they undergo wave-like oscillations (Fig. \ref{fig:knotTrk}B). We now can proceed to quantify multiple such contours as seen in the time-projected contours based on experimental data inputs (Fig. \ref{fig:Analysis}A). These can be projected through the local tangent angle ($\Psi$) dynamics along the filament length as represented in terms of a kymograph (Fig. \ref{fig:Analysis}B) and the distance between the filament ends ($d_e$) demonstrates a compaction over time (Fig. \ref{fig:Analysis}C). 
%

These results demonstrate optimal tracking of intersecting filaments by KnotResolver using simulated and experimental data can be used to study the geometric properties of such curved filaments.


\section*{Discussion}
Here, we have demonstrated the limitations of existing methods and softwares to detect and track self-intersecting filaments in microscopy time-series. We describe a computational image-processing pipeline, KnotResolver, based on intersection detection, mapping to directed graphs, contour tracing in two steps and distance minimization to a template. We demonstrate that this approach results in sub-pixel position detection and high Dice scores in identification, robust to image noise. In comparison to existing tools we demonstrates KnotResolver overcomes their limitations and when applied to experimental fluorescence microscopy time-series images of a `gliding assay' where the filament end is clamped, can allow us to automatically quantify geometry and frequency of oscillations.

Similar to many such software developed KnotResolver has some parameters that need to be user input (Table \ref{tab:krDetails}). These are optimized for microtubules in a dynein gliding assay with pinned filaments. A further test of the utility of this code could be to analyze time-series in more complex backgrounds such as inside cells where compressive forces result in buckling instabilities seen in MT filaments \citep{Brangwynne:2006aa}. We also expect that actin-based gliding assays with typically bent filament with complex curvature, due to the lower persistence length compared to MTs \citep{gittes1993}, could also be tracked using this tool. Indeed surface-defects in myosin-driven actin gliding assays have been previously reported to also result in qualitatively comparable structures \citep{Bourdieu:1995aa}. We expect the use of KnotResolver to extend to simply microtubule filament analysis.

In conclusion our OpenSource computational image analysis tool that uses a novel directed graph based mapping to resolve self-intersections of cytoskeletal filaments from both simulated and experimental data has the potential for automating many of the more unusual contours seen in experimental systems {\it in vitro} and in cells.



\section*{Materials and methods}
\subsection*{Reconstituting dynein-driven microtubule bending in a modified `gliding assay'}
The filament oscillation assays were performed in a flow chamber made by sandwiching double-backed tape between a slide and a coverslip. The chamber was coated with anti-GFP nanobody and streptavidin in a 1:1 molar ratio, followed by casein to block non-specific attachment. Dynein was allowed to attach by incubating in the chamber followed by washes to remove unbound motors. The biotinylated MTs were added to the chamber and allowed to land followed by washes to remove the unbound MTs (Fig. \ref{fig:clampTrkFiesta}A).  Finally a motility buffer containing anti-fade and 4 mM ATP was added to the chamber and MT movement was recorded. The setup was imaged using a 60x (NA 1.45) oil immersion lens using a motorized fluorescence microscope (Nikon TiE, Nikon Corp. Japan) with 10 s time intervals for 10 to 20 min in a lexan enclosure with temperature maintained at 37$^o$ C (Okolab, Pozzuoli, Italy) as previously described \citep{jain2019,Yadav:2024aa}. 
Plus end biotin labelled MTs were prepared by first polymerizing filaments with a 1:4 molar ratio of rhodamine labelled tubulin:unlabelled tubulin, centrifugation to remove monomers and plus labelling with a mixture of 1:4 rhodamine-labelled:biotinylated tubulin, with filaments then stabilized in taxol. 
The filament oscillation assays were performed in double-backed tape flow chambers made by sandwiching double-backed tape between a slide and a coverslip to form a chamber that was coated with the nanobody and streptavidin in a 1:1 molar ratio, followed by casein to block non-specific attachment. To this the anti-GFP nanobody was added, then dynein-GFP, followed by multiple washes to remove unbound motors, biotinylated MTs, more washes and finally motility buffer containing anti-fade and 4 mM ATP. The setup was imaged using a 60x (NA 1.45) oil immersion lens using a motorized fluorescence microscope (Nikon TiE, Nikon Corp. Japan) with 10 s time intervals for 10 to 20 min at 3$^o$C in a temperature controlled Lexan enclosure (Okolab, Pozzuoli, Italy) as previously described \citep{Yadav:2024aa}.

\subsection*{Generating simulated image of oscillatory filaments with increasing `noise'}
{\it Simulations:} We performed simulations of oscillatory wave-like dynamics of simulated MTs based on a stochastic agent-based simulation engine for motor-cytoskeleton interactions, Cytosim \citep{nedelec2007collective}, based on 
a previously described 2D dynamics model of  individual plus-end clamped MTs in a gliding assay \citep{Yadav:2024aa}. 
We systematically varied microtubule lengths from 5 to 50 $\mu m$ and motor density over 12, 25, 50, and 100 motors/$\mu m^{2}$. Each simulation was performed for 5 minutes and 120 frames were saved. The objective was to generate skeleton contours of varying lengths, comparable to experiments. 

{\it Adding `noise' to simulated images:} The simulated skeletons were dilated size-2 diamond-shaped structuring element and convolved the dilated skeletons with a Gaussian point spread function (size: 3, sigma: 2). We introduced noise of varying signal-to-noise ratio ($SNR_{db}$) to the simulated data. The noise model is Gaussian with a specified noise power level $P_{noise}$. The desired $P_{noise }$ is related to $SNR$ by the relation 
\begin{equation}\label{SignalPower}
    P_{noise} = P_{signal}/SNR
\end{equation}
where $P_{signal}$ is equal to the variance of intensity of the original image. $SNR$ in linear scale is related to $SNR_{db}$ by the following relation 
\begin{equation}\label{NoisePower}
    SNR = 10^{\frac{SNR_{db}}{10}} 
\end{equation}
The $SNR_{db}$ values used were 5,10,20 and 30.  The generated  Gaussian noise is computed as $Noise$, 
\begin{equation}\label{gaussianNoise}
 Noise \sim \sigma_{noise} * \mathcal{N}(0, 1)
\end{equation}
where, $\mathcal{N}$ is a normal distribution with zero mean and  the $\sigma_{noise}$ is equal to 
\begin{equation}\label{sigmaNoise}
\sigma_{noise} = \sqrt{P_{noise}}    
\end{equation}
The $Noise$ is added to the original image to get a noisy image with specific $P_{noise}$.

\subsection*{Quantifying filament detection accuracy}
The simulated dataset described previously, was used to create benchmarks to assess the proposed filament tracking method. We quantified detection accuracy using the simulated dataset as a `gold standard' and some typical measures used in image-analysis accuracy using two measures: (a) S\o rensen-Dice Index, SDI for segmentation and (b) Fr\'echet distance to compare two curves.

{\it Dice Index:} The Dice score or S\o rensen-Dice index \citep{dice1945,sorensen1948} takes into consideration TP: True positives, FP: false positives, FN: False negatives and TN: True negatives and combines them as follows:
\begin{equation}
	\text{} = \frac{2 \times \text{\textit{TP}}}{2 \times \text{\textit{TP}} + \text{\textit{FP}} + \text{\textit{FN}}}
	\label{eq:dice}
\end{equation}
True positives are defined as segmented pixel indices that are within a threshold distance (2 pixels) from the ground truth annotations. The choice of the threshold was based on the point spread function (PSF) used to convolve the filament skeletons to resemble experimental images. False positives are segmented pixels which are not within this threshold. Ground truth pixels that were not detected as TP are counted as false negatives. 

{\it Fr\'echet distance:} This metric computes the minimum distance between two curves which is sufficient to traverse their separate paths \citep{frechet1906,frechet1924distance} using a MATLAB implementation of the {\it discrete Fr\'echet distance}. This provides the maximum separation in distance metric between the two curves. The metric is useful for comparing the overlap of two curves while taking into account the proximity of indices between the two curves. For a pair of duplicate curves Fr\'echet distance should be zero, which indicates a perfect overlap and identical ordering of pixel indices between two curves. Fr\'echet distance is computed between the indices of input curve $I(t) = (I_{x}(t),I_{y}(t))$ and the resolved curve $R(t) = (R_{x}(t),R_{y}(t))$ obtained from the KnotResolver pipeline. Here $t$ signifies the orientation or ordering of the two pixel indices in their respective curves. The Fr\'echet distance is computed as follows:
\begin{equation}
    F(I, R) = \inf_{\alpha, \beta} \max_{t} \{\rho(t, \alpha, \beta)\}
\end{equation}
where $\rho(t, \alpha, \beta)$ is the distance function defined as:
\begin{equation}
    \rho(t, \alpha, \beta) = \sqrt{(I_x(\alpha(t)) - R_x(\beta(t)))^2 + (I_y(\alpha(t)) - R_y(\beta(t)))^2}
\end{equation}
and $\alpha$ and $\beta$ are re-parameterizations of the curves $I(t)$ and $R(t)$, respectively.

\subsection*{Analysis of filament contours}
All analysis was performed in MATLAB R2020b (Mathworks Inc., Natick, MA, USA). We calculate the tangent, normal, and curvature vectors for each smoothed contour by using the Frenet-Serret formulas \citep{mate2017frenet}. Smoothing of the x and y coordinates is performed using the Savitzky-Golay filter \citep{savitzky1964smoothing}. The end-to-end distance of the curve is computed as Euclidean distance between the tip and end of each filament. All results are visualized by plotting the curve with color interpolation which indicates the arrangement of contour indices from tip to end. 

\subsection*{Code Availability}
The tool was written in MATLAB2020b (Mathworks, Natick, MA, USA) and uses functions called from the `Signal Processing Toolbox', `Image Processing Toolbox'  `Statistics and Machine Learning Toolbox' and an optional `Parallel Computing Toolbox'. The code for the filament bending analysis has been made OpenSource and and is available at the Github URL \url{https://github.com/CyCelsLab/MTLoopResolve}, together with an example time-series with optimized segmentation parameters. 

\section*{Acknowledgments}
The authors declare no competing financial interests. Prince Sah was involved in the early stages of the project. DK was supported by a fellowship from the Dept. of Biotechnology, Govt of India (DBT/2018/IISER- P/1154). SAY was supported by a fellowship from the Council of Scientific and Industrial Research CSIR (09/936(0261)/2019- EMR-I). The project was supported by funds from the Indo-French Centre for Promotion of Advanced Research (CEFIPRA) grant number 62T5-D and the Dept. of Biotechnology, Government of India grant number BT/PR40262/BTIS/137/38/2022 both awarded to C.A.A.

\section*{Author contributions}
{Dhruv Khatri:} Software, Validation, Investigation, Formal analysis, Writing- Original draft, Visualization {Shivani A. Yadav:} Methodology {Chaitanya A. Athale:} Conceptualization,  Validation, Resources, Writing: Review \& Editing, Supervision, Project administration, Funding acquisition.

\section*{Data Availability Statement}
{\bf Data in a public, open access repository:} The code for the filament bending analysis has been made OpenSource and and is available at the Github URL \url{https://github.com/CyCelsLab/MTLoopResolve}

\bibliographystyle{plainnat}
\bibliography{loopresBib}

\clearpage
\newpage

\section*{Figures}

\clearpage 
\newpage

\thispagestyle{empty} 

\clearpage 
\newpage


\clearpage 
\newpage


\newpage
\clearpage

\clearpage 
\newpage

\clearpage
\newpage

\clearpage 
\newpage

\section*{Supplemental material}
\begin{enumerate}
    \item Supplemental tables
    \item Supplemental figures 
    \item Supplemental movies
\end{enumerate}

\clearpage
\newpage

\subsection*{Supplemental tables}

\renewcommand{\tablename}{Table}
\renewcommand{\thetable}{S\arabic{table}}    
\setcounter{table}{0} 

\begin{table}[ht!]
\centering
\resizebox{\textwidth}{!}{
\begin{tabular}{|c|c|c|cl}
\cline{1-3}
\textbf{Parameter} & \multicolumn{1}{l|}{\textbf{Typical value}} & \textbf{Description}                                                     &  &  \\ \cline{1-3}
segThresh          & 0.46                                  & Threshold for intensity segmentation                                     &  &  \\ \cline{1-3}
segIteration       & 10                                    & Number of iterations for active contour optimization                     &  &  \\ \cline{1-3}
segContract        & 0.4                                   & Contraction bias for active contours                                     &  &  \\ \cline{1-3}
segSmooth          & 0.2                                   & Smoothing factor for active contours &  &  \\ \cline{1-3}
\end{tabular}}

\caption{{\bf Parameters of KnotResolver.} The parameters typically used for microtubule bending and looping image-time series that can be modified by the user for other input image types.}  
    \label{tab:krDetails}
\end{table}

\clearpage
\newpage

\subsection*{Supplemental figures}
\renewcommand{\figurename}{Figure}
\renewcommand{\thefigure}{S\arabic{figure}}    
\setcounter{figure}{0}

\begin{figure}[ht!]
\centering
\includegraphics[width=0.85\linewidth]{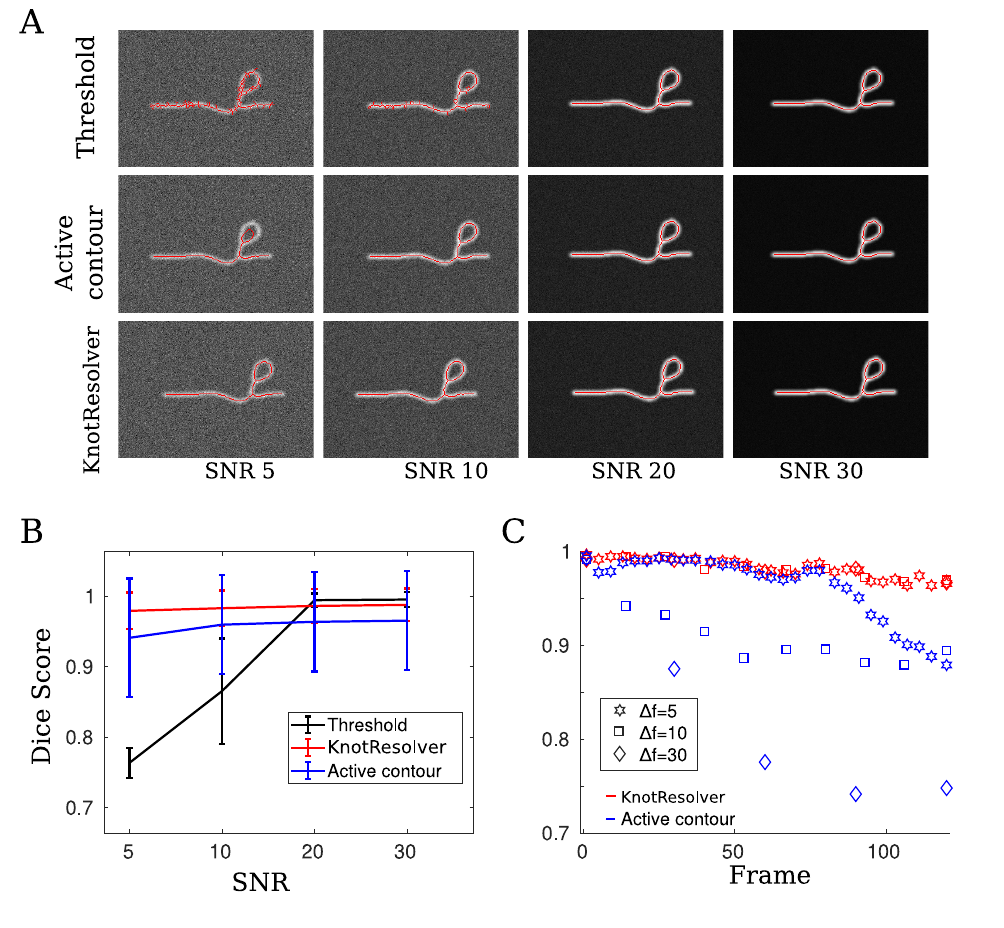}
\caption{\textbf{Dice score of segmentation accuracy by KnotResolver with increasing `noise' compared to simpler approaches.} \textbf{(A)} Montage of segmentation outputs (shown in yellow) overlaid on the input image for the three compared approaches at different SNR. \textbf{(B)} 
Filament segmentation accuracy is compared using the Dice Score (Equation \ref{eq:dice}). Three alternative approaches were tested: threshold-based (black), active contour (blue) and KnotResolver (red). Error bars indicate the standard deviation around the mean. \textbf{(C)} Impact of increasing time step between frames on segmentation accuracy. Segmentation accuracy is compared between KnotResolver (red) and the active contour (blue) method for three frame intervals, $\Delta f$: 5 (\ding{65}), 10 ($\square$), 30 ($\diamond$). Mean Dice scores are computed for every frame across all time series. The analysis involves a total of \textit{n} = 24 time series, each consisting of 120 frames. }
\label{fig:Dice}
\end{figure}

\clearpage 
\newpage

\clearpage
\newpage


\begin{figure}[ht!]
\centering
\includegraphics[width=0.75\linewidth]{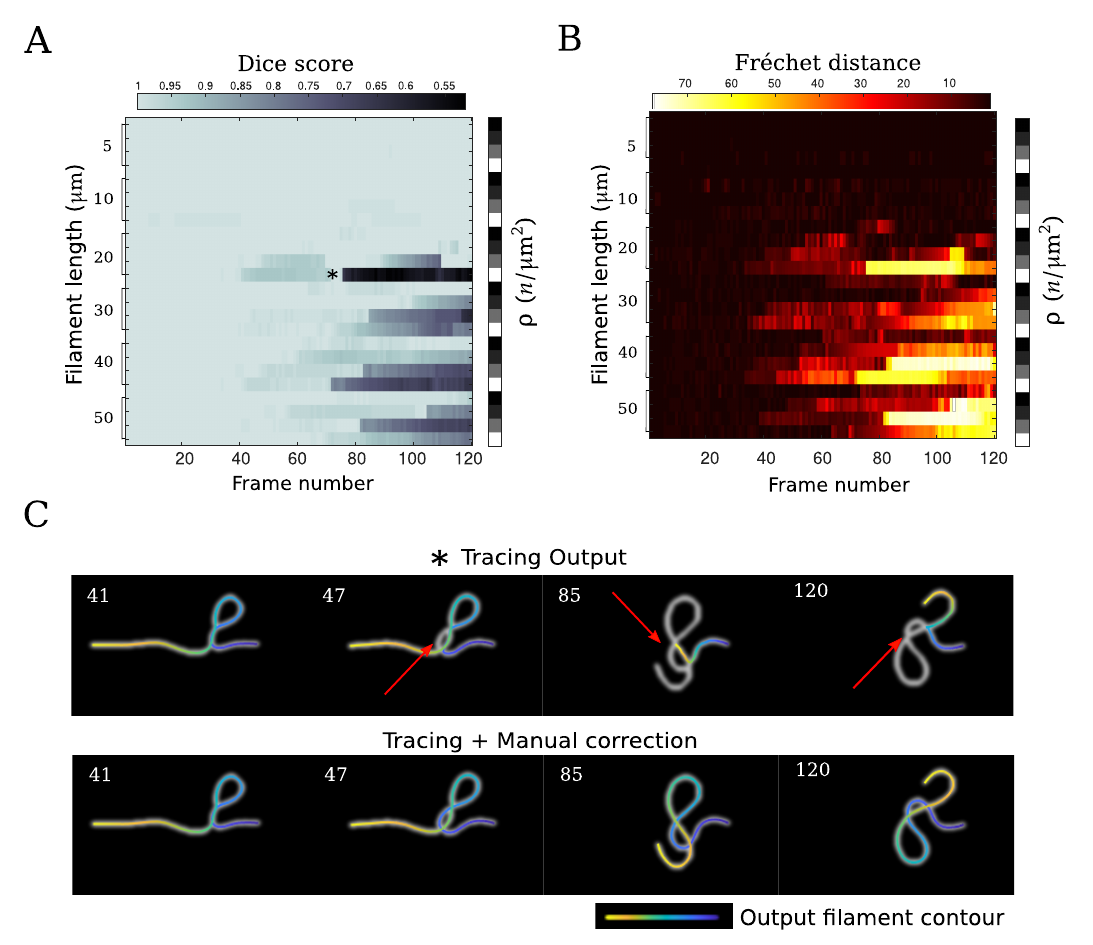}
\caption{\textbf{Interactive resolution of complex knots from simulated data.} (A) Simulated time-series of filament buckling and bending are used as inputs to KnotResolver. Validity of detected contours are quantified in terms of the Dice score (grayscale colorbar). Simulations were run for increasing MT lengths from 5 to 50 $\mu$m (y-axis) and frame-numbers from 1 to 120 (x-axis). Each value of MT length was simulated for four motor densities 12 (black), 25 (dark gray), 50 (light gray) and 100 motors $\mu m^{-2}$ (white). 
Asterisks mark the time series  selected for further analysis. (B) A similar matrix with the Fr\'echet distance is plotted. (C) Branch resolution output using manual restarts at problem areas showing correct arrangement of the filament contour.}
\label{fig:sim}
\end{figure}

\clearpage 
\newpage

\subsection*{Supplemental videos}

\renewcommand{\figurename}{Video}
\renewcommand{\thefigure}{SV\arabic{figure}}    
\setcounter{figure}{0} 


\begin{figure}[ht!]
	\begin{center}
		\includegraphics[width= 0.8\textwidth]{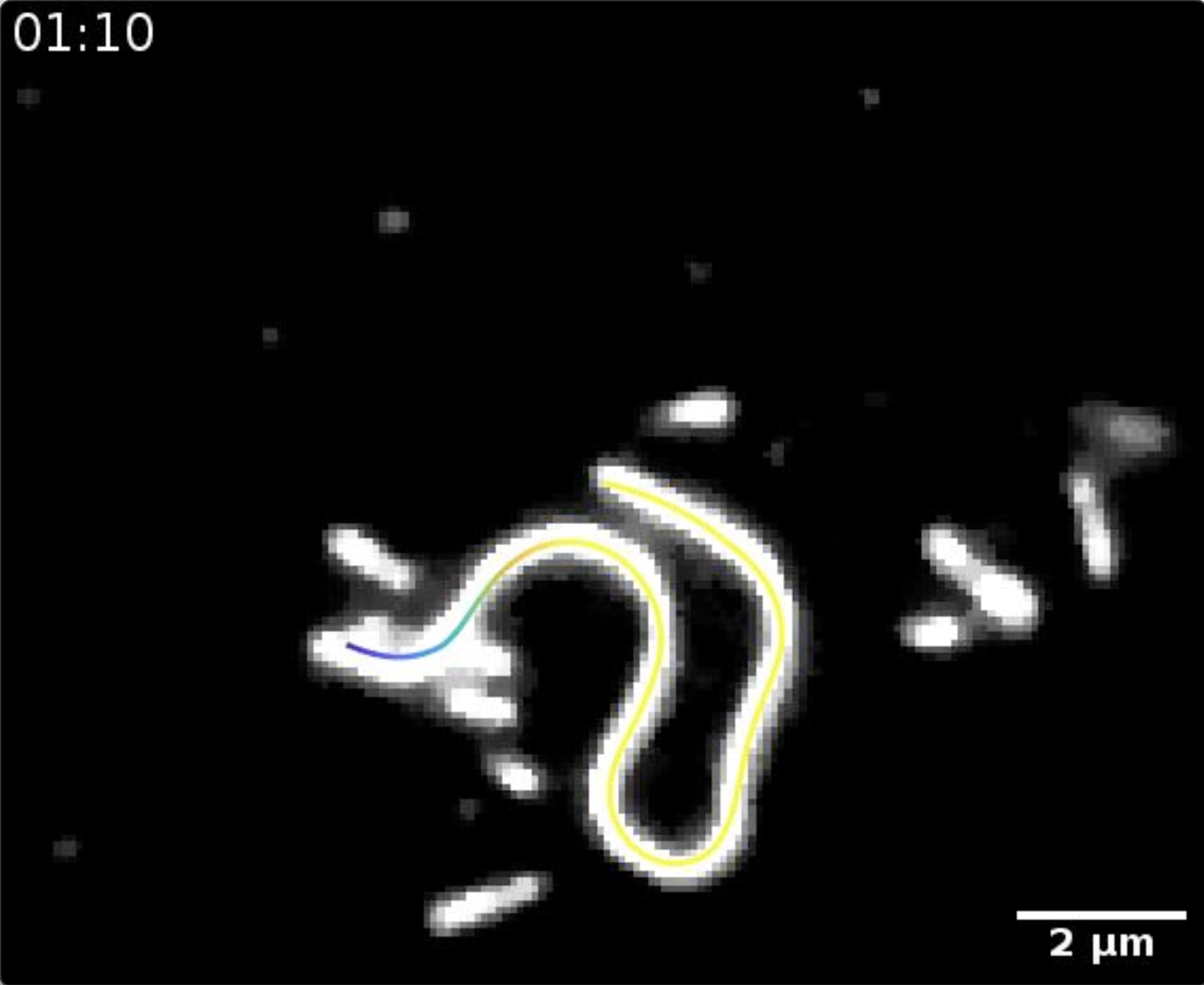}
\end{center}
	\caption{{\bf Representative time-series of filament knot resolution with tracked contour.} The fluorescence microscopy time-series of a filament undergoing self-intersection (gray) overlaid with the segmented and tracked contours. KnotResolver parameters used to automatically track the series without manual intervention were: intensity threshold = 0.46, contour iteration = 10, contraction bias = 0.4, smooth factor = 0.2. Scalebar: 2 $\mu m$, interval between frames ($\Delta t$): 10 s.}
	\label{SV1}
\end{figure}

\clearpage
\newpage

%


\clearpage
\newpage

\end{document}